# Random Bitstream Generation using Voltage-Controlled Magnetic Anisotropy and Spin Orbit Torque Magnetic Tunnel Junctions

Samuel Liu, Jaesuk Kwon, Paul W. Bessler, Suma Cardwell, Catherine Schuman, J. Darby Smith, James B. Aimone, Shashank Misra, and Jean Anne C. Incorvia

*Abstract*— Probabilistic computing using random number generators (RNGs) can leverage the inherent stochasticity of nanodevices for system-level benefits. Device candidates for this application need to produce highly random "coinflips" while also having tunable biasing of the coin. The magnetic tunnel junction (MTJ) has been studied as an RNG due to its thermally-driven magnetization dynamics, often using spin transfer torque (STT) current amplitude to control the random switching of the MTJ free layer magnetization, here called the stochastic write method. There are additional knobs to control the MTJ-RNG, including voltage-controlled magnetic anisotropy (VCMA) and spin orbit torque (SOT), and there is need to systematically study and compared these methods. We build an analytical model of the MTJ to characterize using VCMA and SOT to generate random bit streams. The results show that both methods produce high quality, uniformly distributed bitstreams. Biasing the bitstreams using either STT current or an applied magnetic field shows an approximately sigmoidal distribution vs. bias amplitude for both VCMA and SOT, compared to less sigmoidal for stochastic write for the same conditions. The energy consumption per sample is calculated to be 0.1 pJ (SOT), 1 pJ (stochastic write), and 20 pJ (VCMA), revealing the potential energy benefit of using SOT and showing using VCMA may require higher damping materials. The generated bitstreams are then applied to two example tasks: generating an arbitrary probability distribution using the Hidden Correlation Bernoulli Coins method and using the MTJ-RNGs as stochastic neurons to perform simulated annealing in a Boltzmann machine, where both VCMA and SOT methods show the ability to effectively minimize the system energy with small delay and low energy. These results show the flexibility of the MTJ as a true RNG and elucidate design parameters for optimizing the device operation for applications.

## I. INTRODUCTION

THE magnetic tunnel junction (MTJ) has been widely used in non-volatile random access memory, logic devices, and unconventional computing because of its benefits, e.g., nanosecond speed operation, scalability, thermal stability, high endurance, and compatibility with CMOS [1-5]. More recently, the MTJ is being studied as a probablistic-bit circuit element due to its thermally-driven magnetization dynamics [6-10]. Probabilistic computing leverages the inherent stochasticity of nano-devices, especially to tackle applications in which incorporating stochasticity can provide system-level energy efficiency [11,12], for example when the energy cost of high precision is not needed, or because the application itself is probabilistic and could be more efficiently carried out with stochastic hardware. Probabilistic algorithms require large numbers of random bitstreams, where a stochastic device (e.g. switching between two states) is sampled over time. While a logic device may be evaluated by its switching speed and energy, a probabilistic device is evaluated by the *quality* of random number generated (i.e. is it a true random number?), the *quantity* of random numbers that can be generated per unit time, the *type* of random bitstream that is generated (e.g. is the distribution of random numbers uniform, Gaussian, etc.), and the *controllability* of the random number distribution (i.e. can the probability of measuring 0 or 1 be controlled).

The MTJ has been used as a stochastic-bit circuit element using two main methods: *stochastic read* and *stochastic write*. The MTJ is promising for this application because both of these methods offer controllability, where the probability of generating a 0 or 1 can be biased using an applied current or magnetic field [13]. In the stochastic read method, the MTJ is designed with its free layer (FL) near the superparamagnetic limit, such that it thermally fluctuates in time [14-16]. This produces a high quantity of bitstreams, since the thermal

This work was supported from the DOE Office of Science (ASCR / BES) Microelectronics Co-Design project COINFLIPS and supported from the National Science Foundation Graduate Research Fellowship Program under Grant No. 2021311125 (SL) and supported from the National Science Foundation "Research Experiences for Undergraduates" in accordance with the NSF program solicitation NSF 19-582, under Grant No. 2006753 (PB).

S. Liu and J. Kwon contributed equally to this work. S. Liu, J. Kwon, P. W. Bessler, and J. A. C. Incorvia are now with the Department of Electrical and Computer Engineering, University of Texas at Austin, Austin, TX 78712 USA (e-mail: incorvia@austin.utexas.edu). C. Schuman is with the Department of Electrical Engineering and Computer Science, University of Tennessee, Knoxville, TN 37996 USA.

S. Cardwell, D. Smith, B. Aimone, and S. Misra are with the Sandia National Laboratories, Albuquerque, NM 87123 USA.

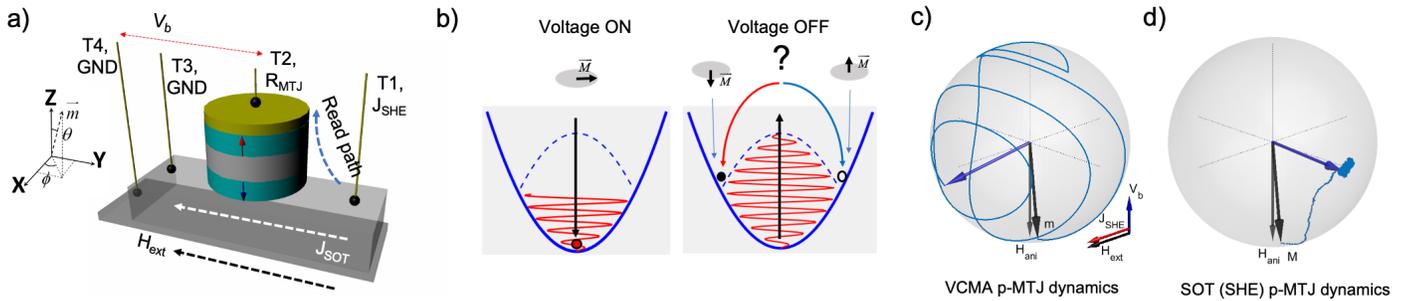

**Figure 1. Schematic and functionality of the magnetic tunnel junction random number generator. a)** Schematic of the MTJ device that can be driven using SOT (current through T1 and T4), VCMA (voltage across T2 and T4), and STT (current across T1 and T2). The ferromagnetic layers are shown in blue. **b)** Schematic of VCMA operation in the form of an energy diagram. The left depicts when voltage is on, and the state is precessing in a middle state, and the right depicts a random settling of the state into one of the two default states. **c)** 3D depiction of VCMA precession to the middle state. **d)** 3D depiction of state transition to the middle state using SOT.

fluctuations are on the order of nanoseconds; but, the bitstream quality is sensitive to temperature, hindering its use. Alternatively, the MTJ can be designed with a thermally-stable FL, and the amplitude of the switching current used to switch the MTJ via spin transfer torque (STT) will determine the switching probability. This stochastic write method is more favorable for scaling to small sizes and robustness against temperature, but could potentially quickly degrade the tunnel barrier due to high current requirements for probabilistic switching in the ballistic regime.

Here, we design an MTJ device that utilizes voltage-controlled magnetic anisotropy (VCMA) and spin orbit torque (SOT) [17-19] to generate random bitstreams. Since micromagnetic solvers are too resource-intensive to generate large numbers of random bitstreams, we build an analytical model of the VCMA/SOT-MTJ that captures the necessary physics, including device geometry, magnetic material parameters, thermal energy, VCMA, STT, and SOT. We show that probabilistic MTJs based on both VCMA and SOT can produce high quality, tunable bitstreams that pass National Institute of Standards and Technology (NIST) uniformity and runs tests for randomness. We compare the energy dissipation and thermal stability of both mechanisms to a stochastic write MTJ. Lastly, we present additional controllability of the VCMA/SOT-MTJs not present in the stochastic read/write MTJs. We show that this control can represent weighted probability problems and also can implement simulated annealing in Boltzmann machines applied to the maximum-satisfiability (MAX-SAT) problem.

## II. DEVICE OPERATION AND THEORETICAL DETAILS

Figure 1a depicts the VCMA/SOT-MTJ device. The MTJ has perpendicular magnetic anisotropy (PMA) and is in a bottom ferromagnetic (FM) FL, top pinned layer (PL) structure [20-24]. The various terminals provide knobs to control the FL switching probability, and in turn the resistance of the MTJ, $R_{MTJ}$, between its two possible resistance states $R_{MTJ} = R_P$ (FM layers parallel *P*) and $R_{MTJ} = R_{AP}$ (FM layers antiparallel *AP*). Terminals T2-T4 (T4 grounded, GND) are used to apply the bias voltage $V_b$ for VCMA. Terminals T1-T2 are used for STT, and terminals T1-T3 for SOT [25-31].

Figures 1b-c depict how the VCMA-MTJ generates random bitstreams through switching into *P* or *AP* states after bias voltage is applied. The PMA free layer starts out with a double-well energy potential (dotted blue line): its magnetization can be *P* (0°) or *AP* (180°) relative to the PL. In Fig. 1b, turning the voltage on from T2-T4 reduces the PMA and sends the FL magnetization in-plane (90°) through precessional motion and damping, depicted by the red lines. Figure 1c shows when the voltage is turned back off, PMA is restored, and the FL transitions into one of two potential *P* or *AP* states, providing the probabilistic bit 1 or 0. For the SOT-MTJ, this mechanism is distinctly different. When a current is applied between T1 and T3 in the heavy metal (HM), SOT generates spin current that is polarized in the in-plane direction. This causes the FL magnetization to be in-plane due to spin current (see Fig. 1d), an effect that does not change the potential energy landscape.

To study the probabilistic behavior of this device, an analytical model is constructed that is physics-based and includes the various knobs to control its switching probability, including thermal energy at room temperature. The magnetization dynamics of the FL is simulated based on a modified Landau-Lifshitz-Gilbert (LLG) equation [32-35],

$$\frac{\partial \vec{m}}{\partial t} = -\gamma \vec{m} \times \vec{H}_{eff} + \alpha \vec{m} \times \frac{\partial \vec{m}}{\partial t}$$
$$- \beta P J_{STT} |\gamma| \vec{m} \times (\vec{m} \times \vec{m}_{pin}) - \beta \eta J_{SOT} \vec{m} \times (\vec{m} \times \vec{\sigma}_{SOT}). \quad (1)$$

where *P* is the spin polarization, $\vec{m}$ and $\vec{m}_{pin}$ are the magnetization of free and pinned layer of the MTJ, respectively, $\gamma$ the gyromagnetic ratio, and $\alpha$ the damping constant. $\beta$ is defined as $\beta = \gamma h / 2 e t_F M_S$ where $e$ is electron charge, $t_F$ is the free layer thickness, and $M_S$ is the saturation magnetization. The first term includes an effective field, $\overrightarrow{H_{eff}}$, and has a uniaxial PMA along $\hat{z}$ with voltage control anisotropy modulations. The second term is the damping term, and the last two terms include current-based magnetization switching. $J_{STT}$ and $J_{SOT}$ are the write current density using STT and SOT; $\overrightarrow{\sigma_{SOT}}$ is the spin current vector from SOT injected into the FL.

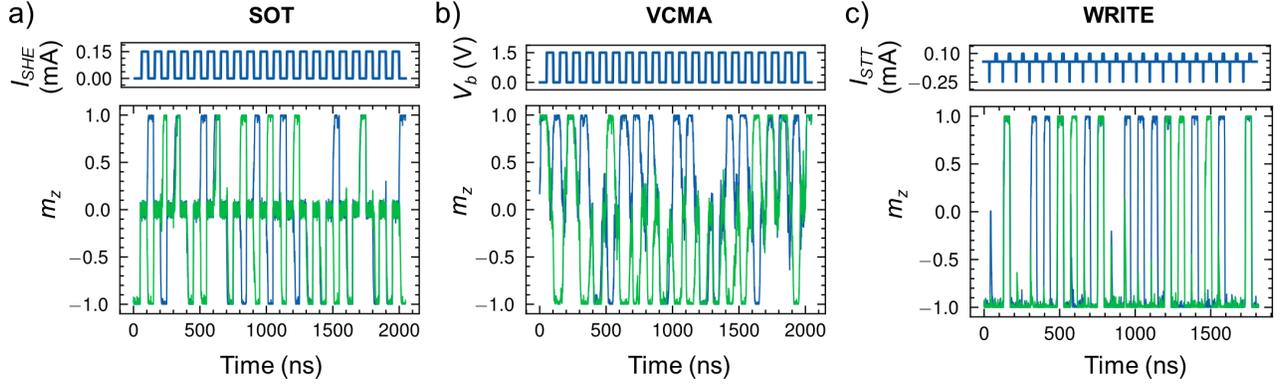

**Figure 2. Stochastic switching dynamics of random number generators.** For **a)** SOT, **b)** VCMA, and **c)** stochastic write MTJ device types, normalized out-of-plane magnetization vector $m_z$ as a function of time is shown for 20 samples over 2000 ns. Two independent runs are shown in green and blue to show stochasticity. The sampling control methods as a function of time are shown in the plots above.

The VCMA effect is included in the LLG equation by an anisotropy field $H_k(V_b)$, which is a function of $V_b$, described as

$$H_k = \frac{(2K_{i,ani})}{(t_{CoFeB} M_S \mu_0)} - \frac{(2K_{s,vcma} V_b)}{(t_{CoFeB} t_{MgO} M_S \mu_0)}. \quad (2)$$

Where $t_{CoFeB}$ and $t_{MgO}$ are the thickness of free layer and tunnel barrier, respectively, $K_{i,ani}$ is the perpendicular magnetic anisotropy (PMA) when $V_b = 0$, and $K_{s,vcma}$ is the VCMA coefficient which describes the interfacial anisotropy between the FL and tunnel barrier under $V_b$.

A thermal fluctuations term is added to perform simulations at non-zero temperature, using [36]

$$\theta_{temp} = \sigma \sqrt{2 \alpha k_b T / (\mu_0 M_S \gamma V \Delta t)}. \quad (3)$$

where $T$ is temperature, $V_{free}$ volume of the FL, and $\sigma$ is a normally-distributed random number. The shape of this distribution is normally-distributed following the noise generated from thermal fluctuations. Following Ref 35 this is input into $\overrightarrow{H_{eff}}$ and is used as a stochastic term every fixed timestep $\Delta t$.

Table 1 lists the parameters and values used in the simulation to represent a standard CoFeB PMA MTJ. The diameter of the circular MTJ is 50 nm, tunnel magnetoresistance (TMR) is chosen to be a reasonable value for PMA MTJs, $TMR = 150\%$, and the FL thickness is chosen to be 1.1 nm. The STT critical switching current ($I_c$) and potential barrier energy ($\Phi_b$) are estimated by

$$I_c = \alpha \frac{\gamma e}{\mu_B g} (\mu_0 M_S) H_k V_{free}. \quad (4)$$

$$\Phi_b = \alpha \frac{\mu_0 M_S H_k V_{free}}{2}. \quad (5)$$

where $V_{free}$ is the volume of FL and $g$ is the spin polarization efficiency factor describes as $g = \sqrt{TMR(TMR+2)} / 2(TMR+1)$. The STT critical switching current density ($J_c = I_c/A$, $A$ is the MTJ area) for the chosen parameters evaluates in the range of $1 \times 10^{11} \sim 4 \times 10^{12} A/m^2$, reasonable for CoFeB [37-42].

## III. STOCHASTIC MTJ BEHAVIOR

Figure 2 shows the bitstream generation using SOT, VCMA,

TABLE I
PHYSICAL PARAMETERS USED IN THE MODEL

| Symbol | Magnetic constant | Values |
|---|---|---|
| $\alpha$ | Gilbert damping | 0.03 |
| $M_S$ | Saturation magnetization | $1.2 \times 10^6$ A/m |
| $H_k$ | Anisotropy effective field | $1.8 \times 10^4$ Oe |
| $A_{ex}$ | Exchange stiffness | $4 \times 10^{-6}$ erg/cm |
| $P$ | Spin polarization of tunnel current | 0.6 |
| $T$ | Temperature | 300 K |
| $k_B$ | Boltzmann constant | $1.38 \times 10^{-23}$ J/K |
| $\mu_0$ | Magnetic permeability | $4\pi \times 10^{-7}$ H/m |
| $e$ | Electron charge | $1.6 \times 10^{-19}$ C |
| Symbol | MTJ | Values |
| $t_{MgO}$ | MgO thickness | $1.5 \times 10^{-9}$ m |
| $t_{free}$ | Free layer thickness | $1.1 \times 10^{-9}$ m |
| $a$ | MTJ diameter | $50 \times 10^{-9}$ m |
| TMR | TMR ration at 0 bias-volt | 150% |
| $\Phi_b$ | MgO potential barrier | ~ 1.2 eV |

and stochastic write. For SOT operation in Fig. 2a, 0.15 mA was selected as the pulse amplitude since it corresponds to approximately 100 mV applied through the bottom HM, typical of low voltage operation of SOT devices. For the VCMA-operated MTJ, 1.5 V applied across the tunnel barrier was chosen to maximize the damping speed while remaining under breakdown voltage. For both, a 30 ns pulse is applied followed by a 30 ns relaxation period. The blue and green series represent two different runs with the same input, showing stochastic bitstream generation represented by $\hat{z}$ magnetization as a function of time. Comparing the output magnetization during pulses, it is evident that the SOT pins the FL magnetization in the in-plane orientation ($m_z = 0$) more strongly than VCMA. This also indicates that in this material system, VCMA is slower to switch than SOT. This is because VCMA relies on a combination of the change in anisotropy as well as damping to switch; more ideal materials for this device type would have a higher damping constant $\alpha$ and VCM coefficient $K_{s,vcma}$.

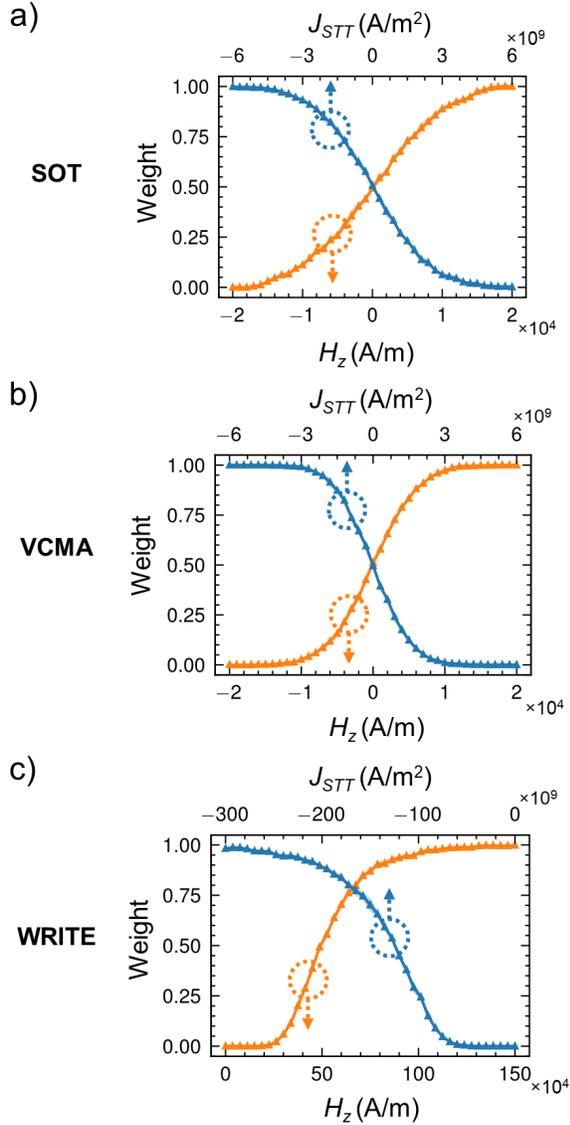

**Figure 3. Probabilistic weight control of stochastic devices.** The probability of sampling a 1 is shown as a function of STT current $J_{STT}$ and magnetic field $H_z$ for **a)** SOT, **b)** VCMA, and **c)** stochastic write MTJ device types, where each point is generated by sampling a given device 2000 times. The weight in response to STT current (magnetic field) is shown in blue (orange). The cloud around each distribution shows the first standard deviation (almost unobservable).

In Fig. 2c, the operation of a stochastic write MTJ is also shown. A 1 ns, -0.25 mA current pulse is used to stochastically write the device. The pulse duration is chosen to ensure that switching occurs in the ballistic regime; the switching probability is relatively independent of temperature. Due to the difference in operation, a 10 ns, opposite polarity current pulse of 10 mA is used to reset the device, with resting periods inserted to match the period of the SOT and VCMA MTJs.

## IV. WEIGHTED BITSTREAM CREATION WITH STT-SOT AND STT-VCMA MTJS

Figure 3 compares control of weighted bitstreams generated with SOT, VCMA, and stochastic write methods. For SOT and VCMA in Figs. 3a and 3b, respectively, the STT current density $J_{STT}$ and magnetic field $H_z$ describe a constant bias applied to the device during operation. The pulse and relax time durations are chosen to be the minimum possible while still producing a high-quality distribution. The pulse (relax) times chosen for the SOT and VCMA MTJs, respectively, are 15 ns (20 ns) and 30 ns (20 ns). For the stochastic write MTJ distribution in Fig. 3c, the $x$ axes represent the current density or magnetic field applied during the 1 ns write pulse. A 10 ns pulse of $5 \times 10^{10}$ A/m² is used to reset the device. 10 ns is chosen to reduce the reset current density in the interest of preserving the tunnel junction. We analyze the average weight $P$ over many cycles using

$$P = \frac{1}{q}\sum_{i=1}^{q} Z(m_z). \quad (6)$$

where $m_z$ is the unit magnetization vector in the $z$ direction, $Z(m_z)$ is the output bitstream function that is 0 when $m_z$ is negative and 1 when $m_z$ is positive, and $q$ is the number of samples; here $q = 2000$ samples, repeated for each point while varying current density and magnetic field separately. The first standard deviation is also shown in the colored cloud around each series, which is almost un-observable. This indicates that for the number of samples chosen, a high quality and accurate stochastic weight can be obtained.

Comparing the distributions between the three device types, the current density required to weight the devices is similar in magnitude for the VCMA and SOT-based devices. Within the material system simulated, the stochastic write MTJ requires currents that are around 500 times larger to stochastically switch the device within 1 ns; this would require larger transistors that can drive larger currents. Comparing the distribution shapes, the SOT and VCMA-based MTJ distributions are approximately sigmoidal, in contrast to that of the stochastic write MTJ. The VCMA/SOT-based MTJs are centered with a weight $P = 0.5$ at zero bias. In an experimental system, this may not be observed due to the presence of stray fields, but this can potentially be alleviated using stack engineering. This centering of $P = 0.5$ at zero bias can be advantageous when producing a uniformly distributed true random number, shown in the subsequent section.

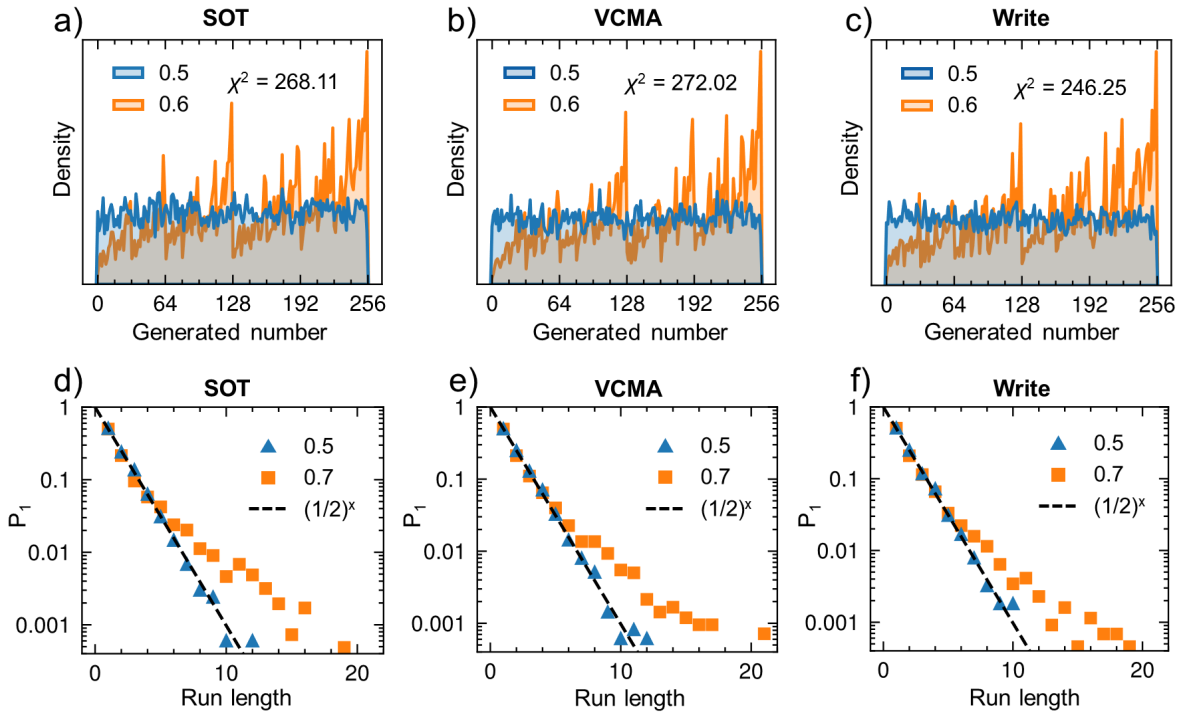

**Figure 4. Bitstream generation quality.** A uniformity test of generated 8-bit numbers is shown in **a-c)** for the three device types, where a $\chi^2$ value is calculated for each distribution weighted to 0.5 (blue). The same test is also applied for devices weighted to 0.6 (orange) to illustrate a distribution generated from a biased device. A runs test recording the probability of observing a given run length of ones in **d-f)** was performed for a sequence of 10000 samples, showing that each subsequent sample is independent for devices weighted to 0.5 (blue), matching the fitted dashed black line of $P_1 = \left(\frac{1}{2}\right)^x$. Results from devices weighted to 0.7 (orange) are shown for comparison.

## V. RANDOM BITSTREAM QUALITY

An immediately attractive application for devices that can stochastically output random bits is as a building block for true random number generators (TRNGs). TRNGs have use in a wide range of applications from neuromorphic computing to hardware security. Therefore, it is important for these random bitstream generators to output high quality random numbers.

Figure 4a-c shows a kernel density estimation of a distribution of random numbers generated to perform the NIST uniformity test for quality of random numbers. For each of the device types, they are weighted to $P = 0.5$ and sampled 8 times continuously to produce an 8-bit number, repeated to produce 10,000 samples. This is also done for $P = 0.6$ to compare with the result from a purposely biased generator. From the plots, all three types of devices can produce any number between 0-255 at an approximately equal probability for $P = 0.5$. The quality of the distribution is evaluated by first computing a normalized $\chi^2$:

$$\chi^2 = \sum_{i=1}^{N} \frac{(O_i - E)^2}{E} \quad (6)$$

where $N$ is the number of possible output numbers (number of bins), $O_i$ is the observed frequency within a given bin, and $E = N/q$ is the expected frequency, where $q$ is the total number of samples. The $\chi^2$ value is a normalized sum of squared deviations from an expected distribution. If the value is small enough, then the observed quantity deviates very little from the expected distribution. For VCMA, SOT, and stochastic write MTJs, the computed $\chi^2$ values are respectively 272.02, 268.11, and 246.25 for a system with 255 degrees of freedom. This corresponds to p-values of 0.2215, 0.2742, and 0.6415 for a $\chi^2$ goodness-of-fit test. These p-values are much larger than 0.01, indicating that all three devices can produce bitstreams that do not significantly deviate from the uniform distribution.

This is further corroborated in the runs tests shown in Figs. 3d-f. This test checks for the length of repeated instances of 1's within a set of continuous samples, chosen to be 10,000. The blue triangles depict the proportion of instances as a function of run length for devices biased to $P = 0.5$. In all cases, the distribution matches the description $P_1 = (1/2)^x$, where $x$ is the length of the run of 1's. Distributions are also presented for devices biased to $P = 0.7$, showing deviation away from the expected distribution. The tests show that all three devices produce bitstreams that are identically distributed, even when weighted.

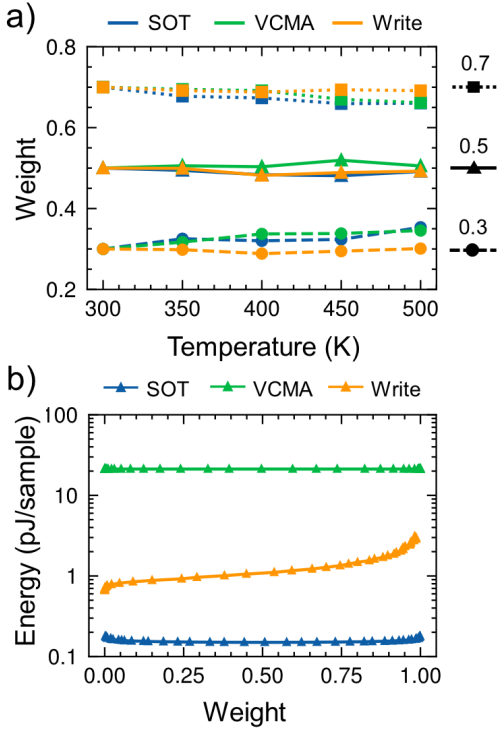

**Figure 5. Temperature dependence and energy dissipation device characteristics. a)** The temperature-dependent probabilistic weighting drift of the of SOT (blue), VCMA (green), and stochastic write (orange) device types are shown for weight levels of 0.3, 0.5, and 0.7. **b)** Intrinsic per sample energy dissipation of each device type as a function of controlled weight.

## VI. Temperature and Energy Comparisons

Because all three device types rely on thermal fluctuations as the source of stochasticity, the effect of temperature changes on biasing of the device is a concern. Figure 5a shows the weight stability of all three types of devices biased to levels of 0.3, 0.5, and 0.7, with 500 samples per point. For all three devices, there is essentially no temperature dependence for a biasing of 0.5 even up to 500 K, though it is important to note that the deviation from 0.5 for the VCMA-MTJ could indicate that 500 samples is not enough to generate a well-calibrated weight. At biases of 0.3 and 0.7, while the apparent weight of the stochastic MTJ remains close to the original at 300 K, there is a clear trend toward 0.5 for the VCMA and SOT-based MTJs. This is because as the thermal fluctuations get larger, a larger current is required to bias the device past the range of the thermal fluctuations during the pulse that sets the VCMA and SOT-based MTJs into the in-plane superposition. The stochastic write MTJ does not experience much of a change due to the operation of the device in the ballistic regime (pulse duration of 1 ns). For thermally-assisted switching of the stochastic write device (pulse duration of 5 ns), a clear trend of increasing bias as temperature increases is shown in Supplemental Fig. S1, where a 5 ns pulse duration is used instead.

For the material system simulated, the energy dissipation is calculated as follows:

$$E = \int_{t_1}^{t_2} \left( \frac{V_b^2}{R_{MTJ}} + I_{STT}^2 R_{MTJ} + I_{SOT}^2 R_{HM} \right) * dt \qquad (7)$$

where $R_{MTJ}$ and $R_{HM}$ are the parallel resistance of the MTJ and resistance of the heavy metal layer, respectively, $I_{SOT}$ is the spin orbit torque current applied through the heavy metal, $I_{STT}$ is the current applied through the junction, and $t_1$ and $t_2$ are the beginning and end times of one sample. Figure 5b shows the energy dissipation necessary to produce one sample for the three device types as a function of the weight. Due to the contrasting physical methods of operation, the energy dissipation between the three device types are on different orders of magnitude. The VCMA-MTJ dissipates on the order of 20 pJ/sample, comparatively higher than that of the stochastic write MTJ at around 1 pJ/sample and the SOT-MTJ at around 0.1 pJ/sample. This is mostly because of the relatively slow damping of the magnetization toward the in-plane superposition, requiring a relatively high-voltage pulse of $V_b = 1.5\ V$ applied for 30 ns, compared to the short write (1 ns) and reset (10 ns) pulses of the stochastic write MTJ and low-current, 15 ns pulse of the SOT-MTJ. As a result, for future development of VCMA-MTJs in this application, materials with high damping $\alpha$ or VCMA coefficient $K_{s,vcma}$ can reduce the energy cost and increase sampling speed.

## VII. Distribution Generator Application

While probabilistic bit streams can be used for a variety of applications, one widely applicable function is to use random number generators (RNGs) to represent a given distribution, which all three device types described can feasibly accomplish. Suppose the distribution we would like to sample from is that of a four-sided die that rolls 0 with probability 1/2, and rolls 1, 2, and 3 with probability 1/6 each. These outcomes can be identified with the outcomes of two coinflips that can be heads (H) or tails (T): 0 corresponds to TT, 1 to TH, 2 to HT, and 3 to HH. If $p$ is the probability of H for coin 1 and $q$ is the probability of H for coin 2, then to represent the desired probability distribution, the following needs to hold:

$$pq = \frac{1}{2} \qquad (8)$$

$$p(1-q) = \frac{1}{6}$$

$$(1-p)(q) = \frac{1}{6}$$

$$(1-p)(1-q) = \frac{1}{6}$$

This is an over-constrained set of equations, rendering no solution for $p$ and $q$, i.e. we do not yet know how to bias the probability of H:T of our device from 50:50 to something else to achieve this probability distribution.

To obtain the desired distribution, we employ a hidden process that induces dependence between observable coins. That is, we assume there is some hidden process that alters the coin's bias, such that for a fraction $r$ of the time coin P has probability of heads $p_1$, and coin Q has probability of heads $q_1$. For the remainder of the time the coins are sampled, the coins have probabilities of heads $p_2$ and $q_2$, respectively. We term this method as the hidden dependence method. This yields the following system of equations that can now be solved:

$$rp_1q_1 + (1-r)p_2q_2 = \frac{1}{2} \tag{9}$$

$$rp_1(1-q_1) + (1-r)p_2(1-q_2) = \frac{1}{6}$$

$$r(1-p_1)q_1 + (1-r)(1-p_2)q_2 = \frac{1}{6}$$

$$r(1-p_1)(1-q_1) + (1-r)(1-p_2)(1-q_2) = \frac{1}{6}$$

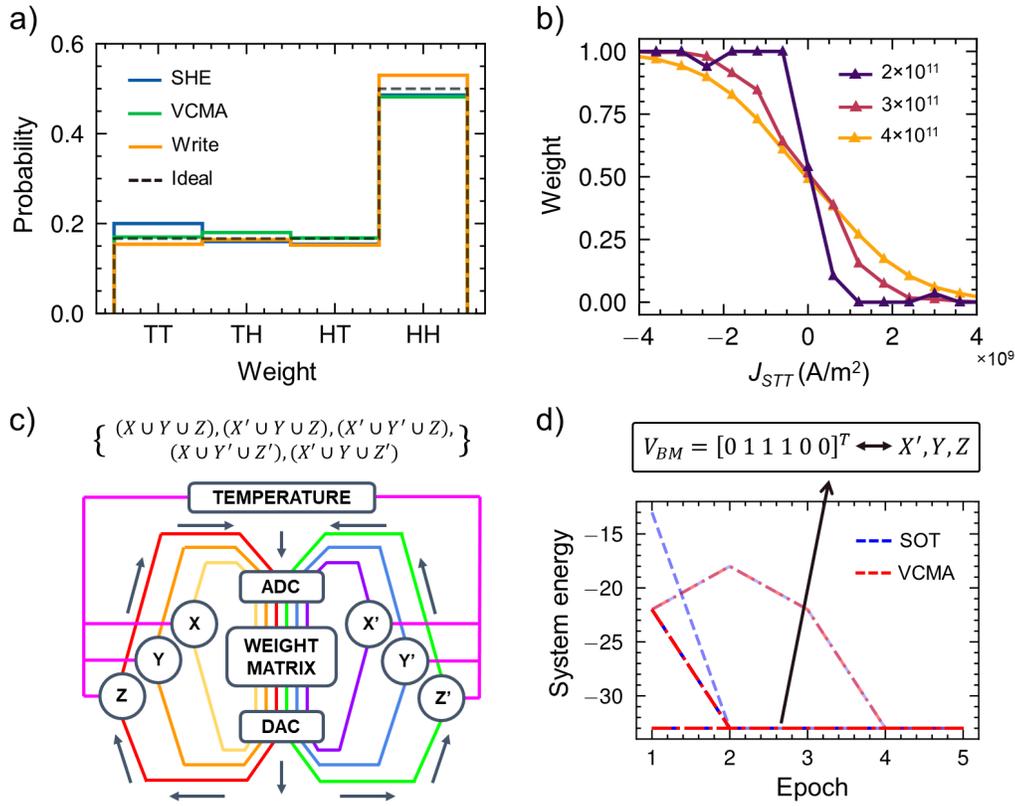

**Figure 6. Stochastic bitstream applications in distribution representation and simulated annealing. a)** Two devices of each device type are used to construct an arbitrary distribution calculated from the Hidden Correlation Bernoulli Coins method, showing a good match with the ideal distribution (black dashed line). **b)** Change in effective temperature by modulating sampling pulse amplitude for an SOT device. Each point is the average of 500 samples. **c)** Schematic of a Boltzmann machine for simulated annealing. The clauses selected for the MAX-SAT test is also shown. **d)** Simulated annealing process of Boltzmann machines constructed from both SOT and VCMA devices for 100 cycles each. The correct solution is also shown in the box above.

The Library of Evolutionary Algorithms in Python (LEAP)[i] is used to solve the equations and determine $r, p_1, q_1, p_2, q_2$ to within a desired tolerance: in this example, it comes out to requiring $r = 0.64925303$, $p_1 = 0.8326947$, $q_1 = 0.84802849$, $p_2 = 0.3542616$, and $q_2 = 0.32530298$. The precision of the quantities depends on the precision needed of the probability distribution. To map out the desired probability distribution, we sample two devices for each device type to represent coin P and coin Q, each 500 times. Each time, when $0 \leq r < 0.64925303$ we bias coin P to $p_1 = 0.8326947$ and then flip it to see if we achieve H or T; similarly, we bias coin Q to $q_1 = 0.84802849$ and then flip it to see if we achieve H or T. When $0.64925303 \leq r \leq 1$, we bias coin P to $p_2 = 0.3542616$ and then flip it to see if we achieve H or T; and, we bias coin Q to $q_2 = 0.32530298$ and then flip it to see if we achieve H or T. Figure 6a shows the resulting probability density of TT, TH, HT, and HH over the 500 coinflips for each of the three devices, showing we approximately achieve the desired probability distribution. We see that there is some deviation from the exact probability distribution. The total energy for flipping the two VCMA, SOT, and stochastic write MTJ coins 500 times each is respectively 13.38 nJ, 0.15 nJ, and 1.37 nJ, corresponding to 25 µs, 12.5 µs, and 10.5 µs total runtime, not including read time. This example shows the usefulness of the stochastic MTJ as a probabilistic device that can produce arbitrary distributions with electrically controllable weighting of the probability.

VIII. SIMULATED ANNEALING APPLICATION

The ability to control the VCMA and SOT MTJ device distributions allow their use as neurons in Boltzmann machines for simulated annealing strategies to solve non-deterministic polynomial-time hard (NP-hard) optimization problems. The idea behind this is that an optimization problem can be translated into a weight matrix, where the solutions to the problem can be solved by minimizing the energy of the system. This is accomplished by using tunable stochastic bit generators as neurons to perform recursive matrix-vector with the set weights, eventually settling into preferred neuron states that depict the solution. Since an optimization problem can have many variables, stochasticity mediated by "temperature" is utilized to check many possible answers by sampling repeatedly. This "temperature" is gradually adjusted in a process known as simulated annealing to obtain an optimized result. Therefore, devices that can produce controllable distributions that emulate temperature effects are good device candidates.

The magnitude of the sampling pulse can be used to adjust the effective temperature of the system, shown in Fig. 6b for the SOT-MTJ. A higher amplitude SOT current pulse ($4\times10^{11}$ A/m$^2$) increases the stochasticity across the applied STT current range, thus resulting in a higher "temperature" while a lower amplitude pulse ($2\times10^{11}$ $A/m^2$) results in a lower "temperature". Analogously, the VCMA-MTJ displays the same behavior with changes in voltage amplitude.

The system-level diagram of the Boltzmann machine for simulated annealing along with a presented optimization problem is shown in Fig. 6c. The chosen problem is a max satisfiability (MAX-SAT) problem that attempts to find a combination of $X$, $Y$, $Z$ Boolean variables that can satisfy the largest number of clauses. The system consists of six neurons $V_{BM}$ to represent the three variables and their complements, along with a 6×6 weight matrix $W_{BM}$. The clauses are translated into weight matrix values as described by Ref. [43]. The system energy is calculated using $E_{BM} = V_{BM}^T W_{BM} V_{BM}$ to visualize the annealing process in Fig. 6d.

As shown in Fig. 6d, only a small number of epochs are necessary to converge to the optimized solution, though the annealing strategy will differ according to the problem. In this case, for both SOT and VCMA, the system temperature is initially set to be high and is decreased to the minimum value in equal increments over 5 epochs. For SOT, the initial maximum sampling current density is set to be $5\times10^{11}$ A/m$^2$ and is decreased to a minimum of $1\times10^{11}$ $A/m^2$. For VCMA, the initial maximum sampling voltage is set to 1.5 $V$ and is decreased to a minimum of 1.1 $V$. Figure 6d shows the result of 100 iterations each of the annealing process, resulting in 100% accuracy in identifying the minimum system energy solution of $X', Y, Z$ at $E_{BM} = -33$. The darker lines depict multiple overlapping iterations. Each iteration took an average energy dissipation of 18.5 pJ for the SOT-MTJ and 305.3 *pJ* for the VCMA-MTJ. The combination of tunability and low energy indicates that both MTJ types can be used effectively as neurons for simulated annealing implementations.

## IX. Conclusion

In summary, we have presented comprehensive simulations of an MTJ-based device that can utilize either VCMA or SOT along with thermal fluctuations to generate high-quality bitstreams, enabling their use as TRNGs. We show that the output bitstream probability can be controlled using STT current or external magnetic field and compare the controllability with a stochastic write MTJ. The quality of output bitstreams using all three device types is evaluated on the NIST uniformity and runs tests, with all three devices performing past the threshold necessary for a TRNG. The energy dissipation, delay, and thermal stability are compared between the device types. With the presented model parameters, the VCMA-MTJ has the largest delay and energy dissipation, and we evaluate that material optimizations for higher damping $\alpha$ and higher VCMA coefficient $K_{s,vcma}$ could bring device performance to parity. We then present two example applications for the devices. Firstly, the devices were shown to be able to generate an arbitrary probability distribution. Secondly, the VCMA and SOT MTJs were shown to have a degree of controllability emulating temperature by modulating pulse duration, allowing use as stochastic neurons to perform simulated annealing in a Boltzmann machine. Both device types were shown to effectively minimize the system energy with small delay and low energy. This work proposes and evaluates novel device architectures that demonstrate promise in implementing tunable RNGs that can be the building blocks of stochastic and neuromorphic systems.

Acknowledgments

We acknowledge support from the DOE Office of Science (ASCR / BES) Microelectronics Co-Design project COINFLIPS. We also acknowledge support from the National Science Foundation Graduate Research Fellowship Program under Grant No. 2021311125 (SL) and support from the National Science Foundation "Research Experiences for Undergraduates" in accordance with the NSF program solicitation NSF 19-582, under Grant No. 2006753 (PB).

This paper describes objective technical results and analysis. Any subjective views or opinions that might be expressed in the paper do not necessarily represent the views of the U.S. Department of Energy or the United States Government. Sandia National Laboratories is a multimission laboratory managed and operated by National Technology & Engineering Solutions of Sandia, LLC, a wholly owned subsidiary of Honeywell International Inc., for the U.S. Department of Energy's National Nuclear Security Administration under contract DE-NA0003525.

References

[1] K. Kobayashi, W. A. Borders, S. Kanai, K. Hayakawa, H. Ohno and S. Fukami, "Sigmoidal curves of stochastic magnetic tunnel junctions with perpendicular easy axis," *Appl. Phys. Lett.*, vol. 119, no. 13, pp. 132406-1–132406-5, Sept. 2021.
[2] Y. Takeuchi, E. C. I. Enobio, B. Jinnai, H. Sato, S. Fukami and H. Ohno, "Temperature dependence of intrinsic critical current in perpendicular easy axis CoFeB/ MgO magnetic tunnel junctions," *Appl. Phys. Lett.*, vol. 119, no. 24, pp. 242403-1–242403-5, Dec. 2021.
[3] D. M. Lattery, D. Zhang, J. Zhu, X. Hang, J. P. Wang and X. Wang, "Low Gilbert Damping Constant in Perpendicularly Magnetized W/CoFeB/MgO Films with High Thermal Stability," *Sci. Rep.*, vol. 8, no. 1, pp. 13395-1–13395-9, Sept. 2018.
[4] M. Urdampilleta, D. J. Niegemann, E. Chanrion, B. Jadot, C. Spence, P. A. Mortemousque, et al., "CMOS with a spin," *Nat. Nanotechnol.*, vol. 14, no. 8, pp.737-741, Aug. 2019.

[5] D. Edelstein, M. Rizzolo, D. Sil, A. Dutta, J. DeBrosse, M. Wordeman, et al., "A 14 nm Embedded STT-MRAM CMOS Technology," 2020 IEEE International Electron Device Meeting, Available: DOI: 10.1109/iedm13553.2020.9371922.

[6] K. Hayakawa, S. Kanai, T. Funatsu, J. Igarashi, B. Jinnai, W. A. Borders, et al., "Nanosecond Random Telegraph Noise in In-Plane Magnetic Tunnel Junctions," *Phys. Rev. Lett.,* vol. 126, no. 11, pp. 117202-1–117202-6, Mar. 2021.

[7] W. A. Borders, A. Z. Pervaiz, S. Fukami, K. Y. Camsari, H. Ohno and S. Datta, "Integer factorization using stochastic magnetic tunnel junctions," *Nature*, vol. 573, no. 7774, pp. 390–393, Sept. 2019.

[8] C. Safranski, J. Kaiser, P. Trouilloud, P. HaSOTmi, G. Hu and J. Z. Sun, "Demonstration of Nanosecond Operation in Stochastic Magnetic Tunnel Junctions," *Nano Lett.* vol. 21, no. 5, pp. 2040–2045, Mar. 2021.

[9] K. Y. Camsari, B. M. Sutton and S. Datta, "p-Bits for Probabilistic Spin Logic," *Appl. Phys. Rev.* vol. 6, no. 1, pp. 1931–9401, Mar. 2019.

[10] S. Jung, H. Lee, S. Myung, H. Kim, S. K. Yoon, S. W. Kwon, et al., "A crossbar array of magnetoresistive memory devices for in-memory computing," *Nature*, vol. 601, no. 7892, pp. 211-216, Jan. 2022.

[11] J. Kaiser and S. Datta, "Probabilistic computing with p-bits," *Appl. Phys. Lett.*, vol. 119, no. 15, Oct. 2021.

[12] J. D. Smith, A. J. Hill, L. E. Reeder, B. C. Franke, R. B. Lehoucq, O. Parekh, et al. "Neuromorphic scaling advantages for energy-efficient random walk computations," *Nat. Elec.*, vol. 5, no. 2, Feb. 2022.

[13] E. Grimaldi, V. Krizakova, G. Sala, F. Yasin, S. Couet, G. Sankar Kar, et al., "Single-shot dynamics of spin–orbit torque and spin transfer torque switching in three-terminal magnetic tunnel junctions," *Nat. Nanotechnol*. vol. 15, no. 2, pp. 111-117, Feb. 2020.

[14] A. Pushp, T. Phung, C. Rettner, B. P. Hughes, S. H. Yang, and S. S. Parkin, "Giant thermal spin-torque–assisted magnetic tunnel junction switching," *Proc. Natl. Acad. Sci.*, vol. 112, no. 21, pp. 6585-90, May 2015.

[15] X. Wang, Y. Zheng, H. Xi, and D. Dimitrov, "Thermal fluctuation effects on spin torque induced switching: Mean and variations," *J of Appl. Phys.*, vol. 103, no. 3, pp. 034507-1–034507-4, Feb. 2008.

[16] V. Krizakova, K. Garello, E. Grimaldi, G. S. Kar and P. Gambardella, "Field-free switching of magnetic tunnel junctions driven by spin–orbit torques at subns timescales," *Appl. Phys. Lett.*, vol. 116, no. 23, pp. 232406-1–232406-5, Feb. 2020.

[17] N. Maciel, E. Marques, L. Naviner, H. Cai and J. Yang, "Voltage-Controlled Magnetic Anisotropy MeRAM Bit-Cell over Event Transient Effects," *J. of Low Power Electronics and Applications,* vol. 9, no. 2, pp. 1-15, Apr. 2019.

[18] S. Wang, H. Lee, F. Ebrahimi, P. K. Amiri, K. L. Wang, and P. Gupta, "Comparative Evaluation of Spin-Transfer-Torque and Magnetoelectric Random Access Memory," *IEEE J. on Emerging and Selected Topics in Circuits and Systems*, vol. 6, no. 2, pp. 134-145, June 2016.

[19] X. Li, K. Fitzell, D. Wu, C. T. Karaba, A. Buditama, G. Yu, et al., "Enhancement of voltage-controlled magnetic anisotropy through precise control of Mg insertion thickness at CoFeB|MgO interface," *Appl. Phys. Lett.,* Vol. 110, no. 5, pp. 052401- 052401-5, Jan. 2017.

[20] S. Ikeda, K. Miura, H. Yamamoto, K. Mizunuma, H. D. Gan, M. Endo, et al. "A perpendicular-anisotropy CoFeB–MgO magnetic tunnel junction," *Nat. Mater.*, vol. 9, no. 9, pp. 721-724, Sept. 2010.

[21] J. Grollier, D. Querlioz, K. Y. Camsari, K. Everschor-Sitte, S. Fukami and M. D. Stiles, "Neuromorphic spintronics," *Nat. Electron.*, vol. 3, no. 7, pp. 360-370, July 2020.

[22] D. C. Worledge, G. Hu, D. W. Abraham, J. Z. Sun, P. L. Trouilloud, J. Nowak, et al. "Spin torque switching of perpendicular Ta|CoFeB|MgO-based magnetic tunnel Junctions," *Appl. Phys. Lett.*, vol. 98, no. 2, pp. 022501-1–022501-3, Dec. 2011.

[23] C. L. Platt, N. K. Minor and T. J. Klemmer, "Magnetic and Structural Properties of FeCoB Thin Films," IEEE Transactions on Magnetics., vol. 37, no. 4, pp.2302-2304, July 2001.

[24] J. A. C. Invorvia, S. Siddiqui, S. Dutta, E. R. Evarts, J. Zhang, D. Bono, et al., "Logic circuit prototypes for three-terminal magnetic tunnel junctions with mobile domain walls," *Nat. Commun.*, vol. 7, no. 10275, pp. 10275-1-10275-7, Jan. 2016.

[25] S. F. Hiroaki Honjo, Kunihiko Ishihara, Keizo Kinoshita, Yukihide Tsuji, Ayuka Morioka, Ryusuke Nebashi, Keiichi Tokutome, Noboru Sakimura, Michio Murahata, Sadahiko Miura, Tadahiko Sugibayashi, Naoki Kasai, and Hideo Ohno, "Material Stack Design With High Tolerance to Process-Induced Damage in Domain Wall Motion Device," *IEEE Transactions on Magnetics*, vol. 50, no. 11, pp. 1401904, Nov. 2014.

[26] S. Ikeda, J. Hayakawa, Y. Ashizawa, Y. M. Lee, K. Miura, H. Hasegawa, et al., "Tunnel magnetoresistance of 604% at 300K by suppression of Ta diffusion in CoFeB/MgO/CoFeB pseudo-spin-valves annealed at high temperature," *Appl. Phys. Lett.,* vol. 93, no. 8, pp. 082508-1–082508-3, Aug. 2008.

[27] M. Wang and Y. Jiang, "Compact model of nanometer STT-MTJ device with scale effect," *AIP Advances,* vol. 11, no. 2, pp. 025201-1–082508-5, Feb. 2021.

[28] K. Nishioka, S. Miura, H. Honjo, H. Inoue, T. Watanabe, T. Nasuno, et al., "First Demonstration of 25-nm Quad Interface p-MTJ Device With Low Resistance-Area Product MgO and Ten Years Retention for High Reliable STT-MRAM," *IEEE Transactions on Electron Devices*, vol. 68, no. 6, pp. 2680-2685, June 2021.

[29] D. Das, A. Tulapurkar and B. Muralidharan, "Scaling Projections on Spin-Transfer Torque Magnetic Tunnel Junctions," *IEEE Transaction on Electron Devices*, vol. 65, no. 2, pp. 724-732, Feb. 2018.

[30] S. Fukami, T. Anekawa, C. Zhang and H. Ohno, "A spin–orbit torque switching scheme with collinear magnetic easy axis and current configuration," *Nat. Nanotechnol.* vol. 11, no. 7, pp. 621-625, July 2016.

[31] L. Thomas, G. Jan, J. Zhu, H. Liu, Y.-J. Lee, S. Le, et al., "Perpendicular spin transfer torque magnetic random access memories with high spin torque efficiency and thermal stability for embedded applications," *J. of Appl. Phys.*, vol. 115, no. 17, pp. 172615-1–172615-6, Oct. 2014.

[32] M. J. Donahue and D. G. Porter, (2002), See http://math.nist.gov/oommf/, for OOMMF User's Guide, NIST

[33] Y. Wang, H. Cai, L. A. d. B. Naviner, Y. Zhang, X. Zhao, E. Deng, et al., "Compact Model of Dielectric Breakdown in Spin-Transfer Torque Magnetic Tunnel Junction," *IEEE Transactions on Electron Devices*, vol. 63, no. 4, pp. 1649-1653, Jun. 2016.

[34] D. Wang, C. Nordman, Z. Qian, J. M. Daughton and J. Myers, "Magnetostriction effect of amorphous CoFeB thin films and application in spin-dependent tunnel junctions," *J of Appl. Phys.*, vol. 97, no. 10, pp. 10c906-1–10c906-3, May 2005.

[35] W. G. Wang, M. Li, S. Hageman and C. L. Chien, "Electric-field-assisted switching in magnetic tunnel junctions," *Nat. Mater.*, vol. 11, no. 1, pp. 64-68, Jan. 2012.

[36] J. Leliaert, J. Mulkers, J. De Clercq, A. Coene, M. Dvornik and B. Van Waeyenberge, "Adaptively time stepping the stochastic Landau-Lifshitz-Gilbert equation at nonzero temperature: Implementation and validation in MuMax[3]," *AIP Advances*, vol. 7, no. 12, pp. 125010-1–125010-13 (2017).

[37] J. Hayakawa, S. Ikeda, Y. M. Lee, R. Sasaki, T. Meguro, F. Matsukura, et al., "Current-Driven Magnetization Switching in CoFeB/MgO/CoFeB Magnetic Tunnel Junctions," *Jpn. J. Appl. Phys.*, vol. 44, no. 41, pp. 37-41, Sept. 2005.

[38] Y. Wang, H. Cai, L. A. d. B. Naviner, Y. Zhang, X. Zhao, E. Deng, et al., "Compact Model of Dielectric Breakdown in Spin-Transfer Torque Magnetic Tunnel Junction," *IEEE Transactions on Electron Devices*, vol. 63, no. 4, pp 1762-1767, Apr 2016.

[39] S. Kanai, Y. Nakatani, M. Yamanouchi, S. Ikeda, H. Sato, F. Matsukura, et al., "Magnetization switching in a CoFeB/MgO magnetic tunnel junction by combining spintransfer torque and electric field-effect," *Appl. Phys. Lett.*, vol. 104, no. 21, pp. 212406-1–212406-3, May 2014.

[40] C. Zhang, S. Fukami, H. Sato, F. Matsukura and H. Ohno, "Spin-orbit torque induced magnetization switching in nano-scale Ta/CoFeB/MgO," *Appl. Phys. Lett.*, vol. 107, no. 1, pp. 012401-1–012401-4, July 2015.

[41] R. Ramaswamy, J. M. Lee, K. Cai and H. Yang, "Recent advances in spin-orbit torques: Moving towards device applications," *Appl. Phys. Rev.*, vol. 5, no. 3, pp. 031107-1–031107-18 Sept. 2018.

[42] Z. Wang, W. Zhao, E. Deng, J.-O. Klein and C. Chappert, "Perpendicular-anisotropy magnetic tunnel junction switched by spin-Hall-assisted spin-transfer torque," *J. Phys. D: Appl. Phys.*, vol. 48, no. 6, pp. 065001-1–065001-7, Jan. 2015.

[43] M. N. Bojnordi and E. Ipek, "Memristive Boltzmann machine: A hardware accelerator for combinatorial optimization and deep learning," *2016 IEEE International Symposium on High Performance Computer Architecture (HPCA)*, pp. 1-13, Apr. 2016.